# Thick GEM-like (THGEM) detectors and their possible applications

R. Chechik, M. Cortesi, A. Breskin
*Dept. Particle Physics, Weizmann Institute of Science, Israel*

D. Vartsky, D. Bar
*SOREQ NRC, Yavne, Israel*

V. Dangendorf
*PTB, Braunschweig, Germany*

Thick GEM-like (THGEM) electrodes are robust, high gain gaseous electron multipliers, economically-manufactured by standard drilling and etching of thin printed circuit board or other materials. Their operation and structure are similar to that of standard GEMs but with 5 to 20-fold expanded dimensions. Due to the larger hole dimensions they provide up to $10^5$ and $10^7$ charge multiplication, in a single- and in two-electrode cascade, respectively. The signal rise time is of a few ns and the counting-rate capability approaches 10 MHz/mm$^2$ at $10^4$ gains. Sub-mm localization precision was demonstrated with a simple, delay-line based 2D readout scheme. These multipliers may be produced in a variety of shapes and sizes and can operate in many gases. They may replace the standard GEMs in many applications requiring very large area, robust, flat, thin detectors, with good timing and counting-rate properties and modest localization. The properties of these multipliers are presented in short and possible applications are discussed.

## 1. INTRODUCTION

The thick GEM-like (THGEM) gaseous electron multiplier (figure 1) was first introduced a couple of years ago [1] and its operation was systematically studied [2, 3], both at atmospheric and at low gas pressures. This multiplier, described in detail in [2], is similar to the standard GEM but with dimensions expanded 5 to 20 fold. The thickness $t$, holes' diameter $d$ and holes' distance $a$, may be chosen according to the application and to the gas operation pressure. The THGEM operation parameters, however, do not scale in proportion, because the charge transport and multiplication process are unaffected. Thus, the electron transport into and from the THGEM holes is much more efficient compared to GEM; the multiplication in a single element is 10-100 times higher (reaching $10^5$) and the cascaded operation mode is very efficient, reaching gains of up to $10^7$ with single electrons. The signal rise-time is of a few ns and counting rate capability is almost 10 MHz/mm$^2$ at $10^4$ gain [2]. The energy resolution for 6 keV x-rays (~20% FWHM) is similar to that of GEM. With 1mm distance between the holes, a localization precision better than a mm is obtained. THGEM-based multipliers are particularly suitable for the efficient detection and imaging of single photoelectrons photo-produced on a solid photocathode coupled to the THGEM or directly deposited on its top face [4]. It was demonstrated that single electrons, originating from a gas ionization gap or from a photocathode deposited on the top of the THGEM, are efficiently detected under much more relaxed conditions compared to standard GEM (e.g. 10-100 times lower THGEM gain).

To complete the study of the THGEM basic properties we have recently investigated the localization capability of THGEM-based detectors, their gain homogeneity and their long-term stability; some of the results will be presented here. We further discuss some possible applications of THGEM-based detectors, in which large-area and robustness are important and sub-mm localization is acceptable.

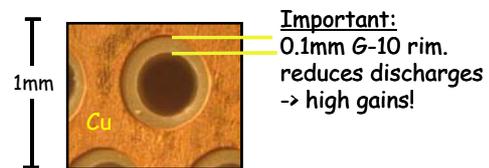

Figure 1: a microscope photograph of a THGEM.

## 2. DETECTORS GAIN AND HOMOGENEITY

The reader is referred to [2] for data on the effective gain of single- and cascaded THGEM elements of various dimensions, in Ar/CH$_4$(95:5) and Ar/CO$_2$(70:30). These data were measured with small THGEM electrodes of typically 3x3cm$^2$ active surface

The gain homogeneity and the localization resolution were studied with a 10x10 cm$^2$ double-THGEM detector (figure 2) irradiated with 8 keV x-rays. It comprised two THGEM electrodes of $t$=0.4mm, $d$=0.5mm and $a$=1mm. They were coupled to a resistive anode (~2MOhm/square) that transmitted the induced signals and broadened them to match the 2mm pitch of the X-Y readout electrode behind it. The latter was a double sided circuit with connected pads, equipped with discrete delay-lines. The detector was operated with Ar/CH$_4$ (95:5) at mean gain $10^4$. The gain variation was of FWHM=10% over the whole surface. The detector's anode and readout electrode are similar to that described in [5].

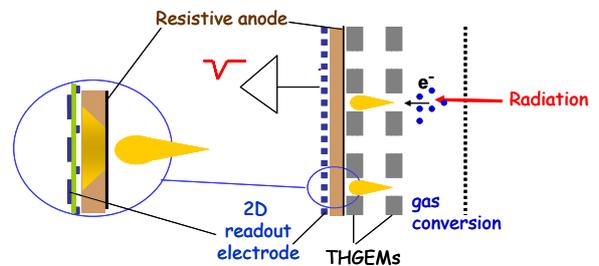

Figure 2. 10x10 cm$^2$ 2D double THGEM detector scheme.

## 3. LONG-TERM GAIN STABILITY

The long-term stability was mostly studied with 3x3cm$^2$ single- and double-THGEM multipliers coupled to a CsI photocathode and illuminated with UV light. It permitted to follow the total current evolution with impinging electron flux of up to $5 \cdot 10^5$ e$^-$/s·mm$^2$, which is equivalent to 1 KHz/mm$^2$ 6 keV x-rays. Some results are shown in figure 3. They indicate at a stabilization time which depends on the history of the electrode (i.e. being the first, second or third measurement cycle); it increases with the total charge, namely the product of the impinging flux and the multiplier gain. The stabilization time is typically 5-12 hours with a THGEM of $t$=0.4mm, $d$=0.3mm and $a$=1mm. It is significantly shorter with $d$=0.6mm. The dependence on the total charge and on the geometry indicates that this is mostly due to charging-up of the insulator surface of the electrode exposed to ions impact. The dependence on the electrode history is still not clear and it might be connected to surface adsorbents such as water.

## 4. 2D IMAGING

The 2D imaging properties were studied with the 10x10cm$^2$ detector described above, irradiated with 8 keV x-rays, and with the help of various masks. Fig 4 shows the raw image accumulated with a standard line-pair mask, while figure 5 gives the modulation transfer function derived from these data. It is obvious that sub-mm localization resolution was obtained with the THGEM of 1mm holes pitch. Recent measurements with an edge mask provided the edge spread function from which a point spread function of FWHM=0.33mm was derived.

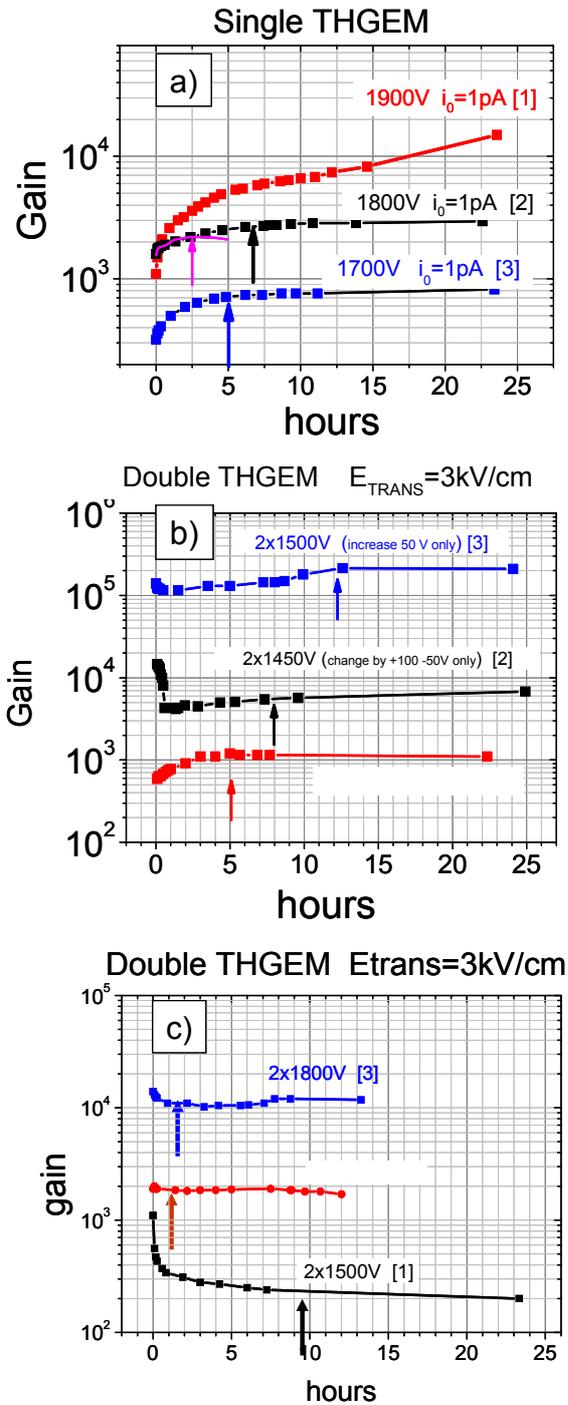

Figure 3. Gain variation measured with a THGEM of $t$=0.4 mm, $d$=0.3 mm and $a$=1mm (a & b) and with a THGEM of $d$=0.6 mm (c), in single (a) and double-element structures (b & c). The numbers on the curves indicate the voltages across each THGEM electrode and the serial measurement number. The stabilization time depends mainly on the total charge and the THGEM geometry, indicating at charging-up of the insulator.

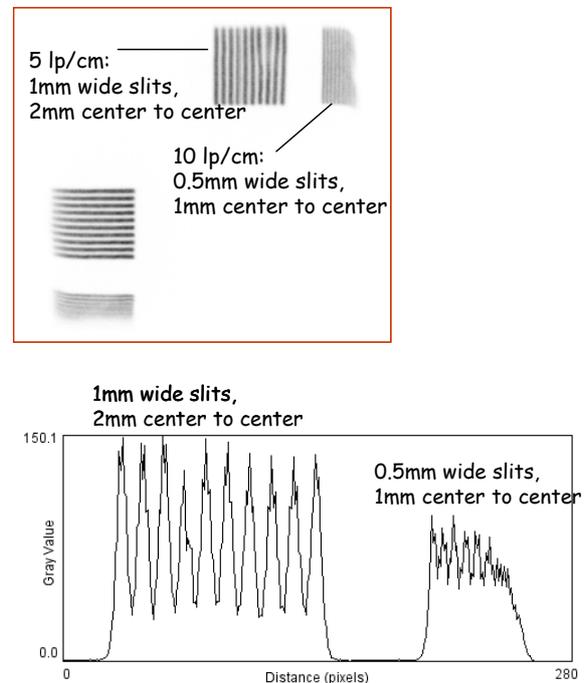

Figure 4. A raw image of part of a line-pair mask (top) and its projection (bottom).

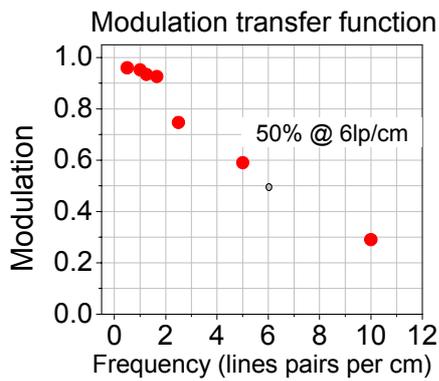

Figure 5. The modulation function derived from a complete line-pair mask, indicating 50% modulation at 6 line-pairs per cm.

The last application could be realized with a reflective photocathode deposited on top of the THGEM; a configuration (figure 6) that ensures high photoelectron detection efficiency and reduced sensitivity to ionizing hadronic background [7].

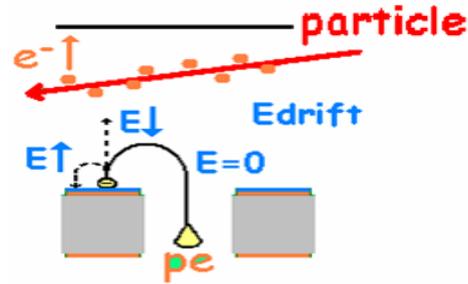

Figure 6. Photon detector incorporating a reflective photocathode coating the THGEM's top surface; with Edrift=0, or with a slightly reversed Edrift, most background ionization electrons are swept away.

## SUMMARY AND POSSIBLE APPLICATIONS

The THGEM multiplier is a robust, low-cost electrode, which may be manufactured with very large area out of a variety of insulating materials with standard (or other) drilling and etching techniques. It provides very large multiplication factors, up to $10^5$ in a single element and $10^7$ in a double-THGEM cascade. These higher gains, compared for example to that of the G-10 "optimized GEM" [6] result from the etched area around the hole's rim. The signals are of a few ns rise time and the rate capability is in the range of a few MHz/mm$^2$. Localization resolution better than a mm (Point spread function 0.33mm) was recorded with 1mm holes pitch and with simple interpolative delay-line readout.

There are several applications in various fields that employ for example GEM electrodes but in fact do not require their very precise localization resolution and can tolerate the 0.4mm thick insulator material of the THGEM. For these applications the THGEM may offer a very attractive alternative solution, at lower cost and better robustness:

- Particle tracking at sub-mm resolutions, such as muon-detectors at the LHC upgrade.
- TPC readout.
- Sampling elements in calorimetry, e.g. for the ILC.
- Moderate-resolution, fast (ns) X-ray and neutron imaging.
- Readout of scintillation light in LXe detectors (e.g. XENON collaboration).
- Single-photon imaging, e.g. Ring Imaging Cherenkov (RICH) detectors (presently done with GEMs for example at the HBD for PHENIX upgrade [7]).

## Acknowledgments

This work was supported in part by the Benoziyo Center for High Energy Research, the Israel Science Foundation, project No151 and by the Binational Science Foundation project No 2002240. A.B. is the W.P. Reuther Professor of Research in the peaceful use of Atomic Energy.